\documentclass[english]{article}
\usepackage[T1]{fontenc}
\usepackage[latin9]{inputenc}
\usepackage{array}
\usepackage{float}
\usepackage{amsmath}
\usepackage{amsthm}
\usepackage{amssymb}
\usepackage{stmaryrd}
\usepackage{graphicx}

\makeatletter

\numberwithin{equation}{section}
\numberwithin{figure}{section}
\theoremstyle{plain}

  \theoremstyle{definition}

\makeatother

\usepackage{babel}
  \providecommand{\definitionname}{Definition}
\providecommand{\theoremname}{Theorem}

\begin{document}

\title{Neural Network for CVA: Learning Future Values}

\author{Jian-Huang She\thanks{Corporate Model Risk, Wells Fargo Bank, email contact: Jian-huang.She@wellsfargo.com}, ~  Dan Grecu\thanks{Corporate Model Risk, Wells Fargo Bank} }

\date{Revised version: Nov. 6, 2018\\
First version: Nov. 6, 2018}
\maketitle
\begin{abstract}
A new challenge to quantitative finance after the recent financial crisis is the study of credit valuation adjustment (CVA), which requires modeling of the future values of a portfolio.  In this paper, following recent work in \cite{Weinan17,Jiequn17},  we apply deep learning to attack this problem. The future values are parameterized by neural networks, and the parameters are then determined through optimization. Two concrete products are studied: Bermudan swaption and Mark-to-Market cross-currency swap. We obtain their expected positive/negative exposures, and further study the resulting functional form of future values. Such an approach represents a new framework for modeling XVA, and it also sheds new lights on other methods like American Monte Carlo.
\end{abstract}

\section{Introduction}

\let\thefootnote\relax\footnotetext{\copyright 2018 Wells Fargo Bank, N. A. All rights reserved. \\The views expressed in this publication are our personal views and do not necessarily reflect the views of Wells Fargo Bank, N.A., its parent company, affiliates and subsidiaries.}

The 2008 financial crisis has brought to central stage counterparty credit risk. Thereafter it has become a crucial task for banks to evaluate the Credit Valuation Adjustment (CVA), or more broadly XVA, including in addition Debit Valuation Adjustment(DVA), Funding Value Adjustment (FVA), Margin Value Adjustment (MVA) etc. The key concept for XVA is the future value of a portfolio. Henceforth, in addition to time-zero price, we now need to model the distribution of prices in the future. Furthermore for collateralized trades, we need to model the joint distribution of prices at different times in the future. For MVA, we need to model derivatives of future values (future greeks). 

In fact, in the (time-zero) pricing of products with early exercise features, e.g. Bermudan swaptions, one already faces the problem of obtaining prices at future time points: the holding values at the exercise dates. This points to the close relationship between XVA and callable products. For callable products, while partial differntial equations (PDE) and tree methods work well in low dimensions, when dimension becomes high, e.g. for Libor Market Model, the only resort is still the Monte Carlo method. 

To draw inspiration, let us examine more closely the use of Monte Carlo method for callable products. It is clear that conventional Monte Carlo method does not work here, since each future value requires a separate Monte Carlo pricer, and such nested Monte Carlo method is too costly. The so-called American Monte Carlo (AMC) is used instead \cite{Longstaff01}. To proceed, one starts with the seemingly obvious observation that future value is determined by the information available till that date. More formally, future value is the conditional expectation value filtered on information available at the corresponding time in the future. The direct consequence of this observation is that future value is a function of the historical time series of all the risk factors on a single Monte Carlo path. Without any barrier or Asian feature, it is further simplified that future value is simply a function of all the risk factors on that particular date and for that particular path. That is, for all the risk factors $X_{i}$ with $i=1,\cdots, d$ , the future value is a function of these d-variables: $V\left(T_{n}\right)=f\left(X_{1}\left(T_{n}\right),\cdots,X_{d}\left(T_{n}\right)\right)$. Note that this function is generally highly nonlinear, and it is defined in high dimensions. 

One way to determine such a function is through AMC, which proceeds with the further observations that (1) given a Monte Carlo path for the whole life of the trade, one can obtain the trade value at a given date for the particular path from discounted cash-flows; (2) the value thus obtained represents a sampling of the unknown function $f\left(X_{1},\cdots,X_{d}\right)$. This way one can obtain a large number of samplings for a single function, and statistical methods can then be used to infer this function. In practice, linear regression is usually used due to its efficiency. The linear regression method assumes fair amount of prior knowledge of the problem at hand, in the form of basis functions. And the performance of this method depends crucially on the intelligent choice of these base functions. For example the prices of the more liquid products that are used to hedge this product are usually included in the set of base functions.

Here we explore the possibility of determining the functional form of the future value without using as much prior knowledge of the products. More concretely we seek to to find a ``universal approximator'' for the high dimensional, non-linear function $f\left(X_{1},\cdots,X_{d}\right)$. We follow closely the seminal work of Wei-Nan E and collaborators  (see \cite{Weinan17,Jiequn17}) to use neural network (NN) as such a ``universal approximator''. This approach is largely inspired by recent success in applying NN to various areas such as computer vision, speech recognition, machine translation, playing board games and medical diagnosis (see \cite{Goodfellow16,Hinton12,Krizhevsky12,LeCun15}).  Two attractive features of the neural network are (1) the power to represent high-dimensional non-linear functions \cite{Cybenko89, Hornik90}, (2) the easiness to determine the involved parameters using efficient optimization algorithms \cite{Hinton86}. 

NN approach has been applied to CVA in \cite{Labordere17} using a dual formulation of stochastic control problems, and to time-zero pricing of Bermudan swaption using Libor-Market Model in \cite{Qi18}. A new algorithm has been proposed in \cite{Raissi18} to directly parameterize the future value with NN, exploring the power of automatic differentiation. 

The rest of this paper is organized as follows. In section 2, we review the basic concepts of CVA/DVA, and then lay out the neural network approach to model future values and compute CVA/DVA. In section 3 and 4, two concrete products, i.e. Bermudan swaptions and Mark-to-Market cross currency swaps, are studied. Their EPE/ENE are computed and the functional form of future values are studied. In section 5, we summarize our approach, and comment on future directions.

\section{The formalism\label{Formalism}}

\subsection{CVA/DVA}

CVA and DVA are the risk-neutral prices of counterparty risk. They are defined as (see e.g. \cite{Brigo13, Gregory15}): 
\begin{eqnarray}
CVA	 &=& \int_{0}^{T}	E_{0}^{Q}\left[\left(1-R_C\left(t\right)\right)D\left(0,t\right)V^{+}\left(t\right)1_{t\le\tau_C<t+dt}\right],\\
DVA &=& \int_{0}^{T}	E_{0}^{Q}\left[\left(1-R_B\left(t\right)\right)D\left(0,t\right)V^{-}\left(t\right)1_{t\le\tau_B<t+dt}\right],
\end{eqnarray}
with discount factor $D\left(0, t\right)$, future value of the portfolio $V\left(t\right)$, recovery rate $R_C$ and $R_B$, default time $\tau_C$ and $\tau_B$ for the counterparty (C) and bank (B). Here $V^{+}\equiv max\left(0,V\right)$, $V^{-}\equiv min\left(0,V\right)$. The indicator functions represent the conditions that counterparty/bank defaults in the time interval $[t, t+dt)$. The expectation is taken under risk-neutral measure conditioned on information at time zero, i.e. valuation time.  

Under the assumption that default events and future prices are independent, CVA/DVA can be separated into the market part and credit part:
\begin{eqnarray}
CVA 	&=&	\int_{0}^{T}\left(1-R_C\left(t\right)\right)EPE\left(t\right) dPr_C\left(t\right),\\
DVA	&=&	\int_{0}^{T}\left(1-R_B\left(t\right)\right)ENE\left(t\right)dPr_B\left(t\right).
\end{eqnarray}
The credit part is encoded in the recovery rates $R_{C/B}$ and default probabilities $dPr_C\left(t\right) \equiv Pr\left(t\le\tau_C <t+dt\right)$, and
 $dPr_B\left(t\right) \equiv Pr\left(t\le\tau_B< t+dt\right)$.
 The market part is encoded in the expected positive exposure (EPE) and expected negative exposure (ENE): 
\begin{eqnarray}
EPE\left(t\right)	&=&	E_{0}^{Q}\left[D\left(0,t\right)V^{+}\left(t\right)\right],
\label{EPE}
\\
ENE\left(t\right)	&=&	E_{0}^{Q}\left[D\left(0,t\right)V^{-}\left(t\right)\right].
\label{ENE}
\end{eqnarray}
Note that EPE represents a call option on the portfolio with strike zero, and ENE a put option. In this paper, we will focus on EPE and ENE, which involve the modeling of future values $V\left(t\right)$.

\subsection{Neural network for future values}

We invoke deep learning to model future values. In this subsection we outline the formulation of the problem. While the approach of \cite{Weinan17,Jiequn17} starts with PDE, and then converts them to backward stochastic differential equations (BSDE \cite{Karoui97, Pardoux90, Pardoux92}), we find it more convenient to start direcly in the framework of BSDE. We start with the risk factor dynamics. We consider $d$ risk factors $X_i$ with $i=1,\cdots,d$ (think for example the Libor forward rates), and $\mathbf{X}_{t}\equiv\left(X_{1}\left(t\right),\cdots,X_{d}\left(t\right)\right)$. Their dynamics reads: 
\begin{equation}
dX_{i}\left(t\right)=\mu_{i}\left(t,\mathbf{X}_{t}\right)dt+\sum_j\sigma_{ij}\left(t,\mathbf{X}_{t}\right)dW_{j}\left(t\right), 
\end{equation}
with the correlation $<dW_{j}\left(t\right),dW_{k}\left(t\right)>=\rho_{jk}dt$, and $j,k=1,\cdots,K$ from a $K$ dimensional Brownian motion.

Consider then the future value $V\left(t\right)$, which is defined as the conditional expectation filtered on information available at time $t$: 
\begin{equation}
V\left(t\right)=B\left(t\right)E\left[\sum_{n}\frac{CF\left(T_{n}\right)}{B\left(T_{n}\right)}|{\cal F}_{t}\right],
\end{equation}
with cashflow $CF$ and numeraire $B$. The natural variable for BSDE of future value is the relative value to the numeraire $B\left(t\right)$, i.e. $\tilde{V}\left(t\right)\equiv V\left(t\right)/B\left(t\right)$, since it is a martingale under the corresponding measure. The resulting dynamics involves no drift but only diffusion: 
\begin{equation}
d\tilde{V}\left(t\right)=\sum_{ij}\frac{\partial\tilde{V}}{\partial X_{i}}\left(t,\mathbf{X}_{t}\right)\sigma_{ij}\left(t,\mathbf{X}_{t}\right)dW_{j}\left(t\right).
\end{equation}
If needed, one can then obtain the BSDE for $V\left( t \right)$ from the above equation.

Of central importance is the unknown vector function $\frac{\partial\tilde{V}}{\partial X_{i}}\left(t,\mathbf{X}_{t}\right)$, which in financial terms represents Delta for the corresponding risk factors. We discretize the time direction. At a given time step $T_{n}$, each Delta is a function of all the risk factors at $T_{n}$: 
\begin{equation}
\frac{\partial\tilde{V}}{\partial X_{i}}\left(T_{n}\right)=f_{i}^{\left(n\right)}\left(X_{1},\cdots,X_{d}\right),
\end{equation}
where the functional form $f_i^{(n)}$ is unkown. The insight of \cite{Weinan17,Jiequn17} is to parameterize this function by a neural network: 
\begin{equation}
\frac{\partial\tilde{V}}{\partial X_{i}}\left(T_{n}\right)\simeq{\cal F}_{i}^{\left(n\right)}\left(X_{1},\cdots,X_{d}\right),
\end{equation}
and then obtain the functional form through optimization.

Consider a fully-connected neural network, which is formed by repeatedly applying two simple operations: 
\begin{itemize}
\item linear combination of all input variables, i.e. $z_{j}=\sum_{i}w_{ji}x_{i}$. 
\item nonlinear mapping of a single variable, i.e. $y_{j}=\phi\left(z_{j}\right)$, where $\phi\left(z\right)$ can be $\tanh\left(z\right) $, $\max\left\{ z,0\right\}$ (relu), $1/\left(1+e^{-z}\right)$ (sigmoid). 
\end{itemize}
We can compare the neural network representation of a function with that of a polynomial representation: the crucial difference is that neural network can have multiple layers. With enough layers, the neural network can essentially represent arbitrarily complicated non-linear functions. The power of this approach really comes from the existence of a fast optimization algorithm \cite{Hinton86}.  

Once the functional forms of the Delta's are given, the evolution of the discounted future values is determined:
\begin{equation}
\tilde{V}\left(T_{n+1}\right)=\tilde{V}\left(T_{n}\right)+\sum_{ij}{\cal F}_{i}^{\left(n\right)}\left(\mathbf{X}_{n}\right)\sigma_{ij}\left(T_{n},\mathbf{X}_{n}\right)\left[W_{j}\left(T_{n+1}\right)-W_{j}\left(T_{n}\right)\right].
\end{equation}
In addition, when there are cashflows or option exercises, the portfolio values can jump. We use $T_{n}^{\pm}$ to represent times immediately after and before date $T_{n}$. At the cashflow dates, one has the jump condition: 
\begin{equation}
V\left(T_{n}^{-}\right)-V\left(T_{n}^{+}\right)=CF\left(T_{n}\right).
\end{equation}
At the option exercise dates, one has the jump condition:
\begin{equation}
V\left(T_{n}^{-}\right)=\max\left\{ V\left(T_{n}^{+}\right),U_{n}\left(T_{n}\right)\right\} ,
\label{eq:V-jump}
\end{equation}
where $U_{n}\left(T_{n}\right)$ denotes the exercise value. Note the asymmetry in time for the option exercise condition: given the exercise value $U_{n}\left(T_{n}\right)$, we can determine $V\left(T_{n}^{-}\right)$ from $V\left(T_{n}^{+}\right)$, but we can not determine $V\left(T_{n}^{+}\right)$ from $V\left(T_{n}^{-}\right)$. Mathematically it is because the max function does not have an inverse function. Financially it is because fair values are determined by expectations of the future. The consequence is that for products involving early exercise, one has to evolve the portfolio value backward in time:
\begin{equation}
\tilde{V}\left(T_{n}\right)=\tilde{V}\left(T_{n+1}\right)-\sum_{ij}{\cal F}_{i}^{\left(n\right)}\left(\mathbf{X}_{n}\right)\sigma_{ij}\left(T_{n},\mathbf{X}_{n}\right)\left[W_{j}\left(T_{n+1}\right)-W_{j}\left(T_{n}\right)\right].
\label{eq:V-backward}
\end{equation}
 This applies to all formalisms including PDE, trees, AMC, and also NN method (see \cite{Qi18}).  

Given the equations governing the evolution of future value, to obtain its full history, we still need the boundary conditions. The final condition is that right after the maturity date $T_{N}$, the portfolio value should be zero: 
\begin{equation}
V\left(T_{N}^{+}\right)=0. 
\end{equation} 
The initial condition is unknown, and will be parameterized and obtained through optimization. We parameterize the initial value $V\left(0\right)=V_{0}$, and the initial Delta's $\frac{\partial V}{\partial X_i}\left(0\right)=Z_i^{(0)}$.

 Given (1) the Monte Carlo paths of the risk factors $X_i\left(T_n, \omega_p\right)$, where $\omega_{p}$ denotes a Monte Carlo path, (2) initial condition for future values, in terms of the parameters $V_0$ and $Z_i^{(0)}$, (3) BSDE for future values, with neural network parameters $w_i^{(n)}$, we can obtain the ``tentative'' history of all the future values $V\left(T_{n},\omega_{p}\right)$.  From such a history, we can construct a loss function that represents the deviation of the involved trial parameters $\left(V_{0},Z_i^{(0)},w_{i}^{\left(n\right)}\right)$ from their ``real'' values. One possible choice for the loss function is the mean squared error of the future values from the target values
\begin{equation}
{\cal L}\left(V_{0},Z_i^{(0)},w_{i}^{\left(n\right)}\right)=\frac{1}{{\cal A}}\sum_{M p}\left[V\left(T_{M},\omega_{p}\right)-V_{\rm target}\left(T_{M}\right) \right]^{2}, 
\end{equation}
with a normalization factor ${\cal A}$, e.g. number of paths. If forward induction is used, we can use the deviation at the maturity date, i.e. $T_M=T_N$, where $V_{\rm target} =0$. If backward induction is used, e.g. with early exercise, we can use the deviation at the valuation date $T_M=T_0$, where $V_{\rm target} =V_0$.

With the constructed loss function, one applies optimization to train the neural network. We use Adam optimization algorithm \cite{Kingma14}, which has been widely adopted for recent deep learning applications in computer vision and natural language processing. 

\subsection{Exposure calculation}

Once the future values are known, the EPE and ENE can be computed from Eq.(\ref{EPE}) and (\ref{ENE}). As we already have the Monte Carlo set up, this step is relatively straightforward for linear products. We evolve the future values one more time with the already trained parameters, and in this process, compute EPE and ENE according to Eq.(\ref{EPE}) and (\ref{ENE}).

 The exposure calculation for options is more involved. We still evolve the future values one more time with the trained parameters, but in this process, we also need to keep track of the exercise time $\tau_p$ for each Monte Carlo path $\omega_p$. If the option has not been exercised, the portfolio value is the value of the option. If the option has been exercised, there are two different cases that need to be treated separately: for cash-settled option, the portfolio value is zero; for physically-settled option, the portfolio value is the value of the underlying. To summarize, one has
 \begin{equation}
V\left( t, X_p \right) = \left\{ \begin{array}{rl}
  V_{\rm option}\left( t, X_p \right) &\mbox{  for $\tau_p> t$,} \\
  0 &\mbox{for  $\tau_p \le t$ and cash-settled,}\\
 V_{\rm underlying}\left( t, X_p \right) &\mbox{for $\tau_p \le t$ and physically-settled.} 
       \end{array} \right.
\end{equation}

\section{Bermudan Swaption}

\begin{figure}[h]
\includegraphics[width=0.49\textwidth]{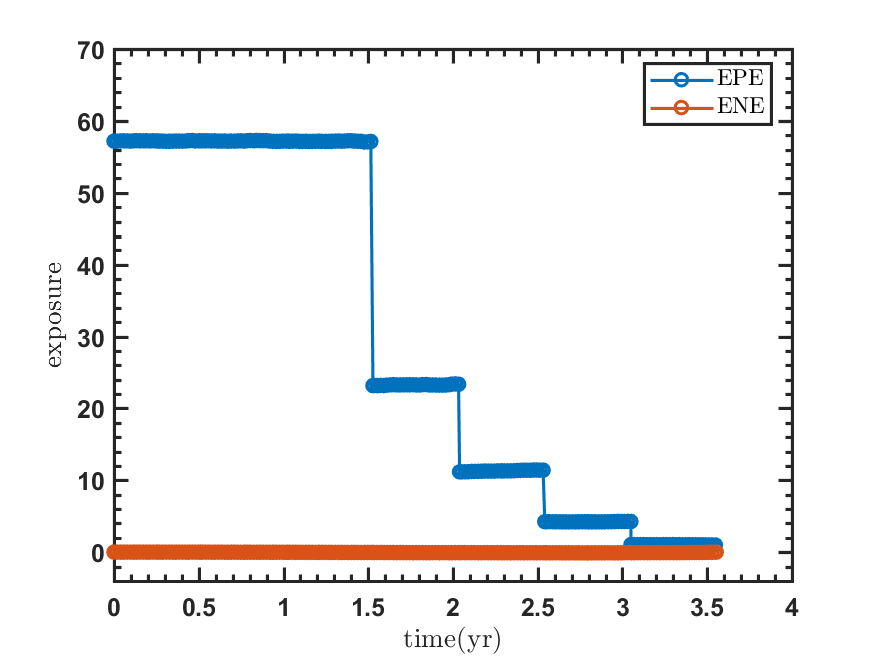}
\includegraphics[width=0.49\textwidth]{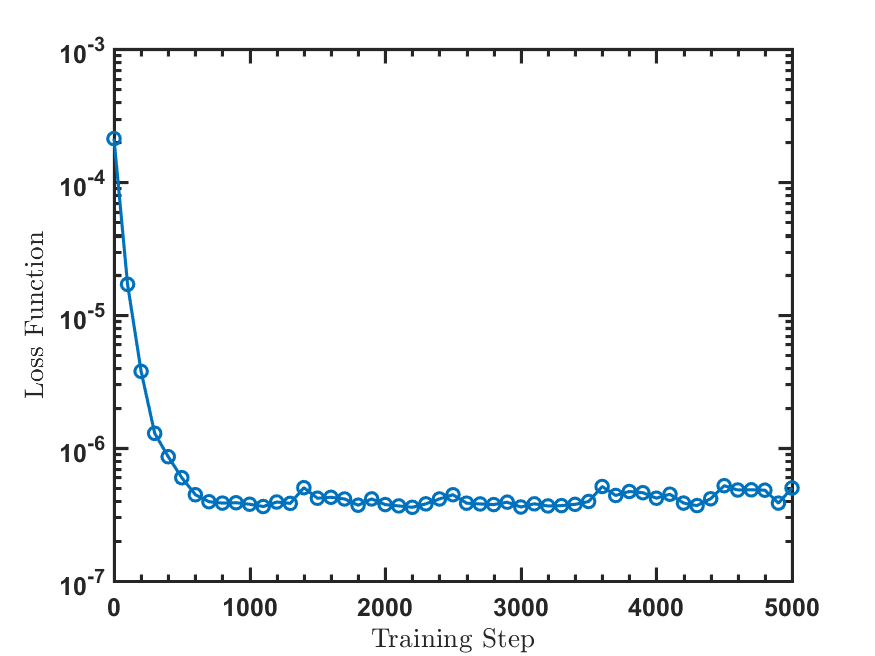}
\caption{Left: EPE and ENE of cash-settled Bermudan swaption. Right: Evolution of loss function with training steps.}
\label{fig:Bermudan-EPE}
\end{figure}

\begin{figure}[h]
\includegraphics[width=0.3\textwidth]{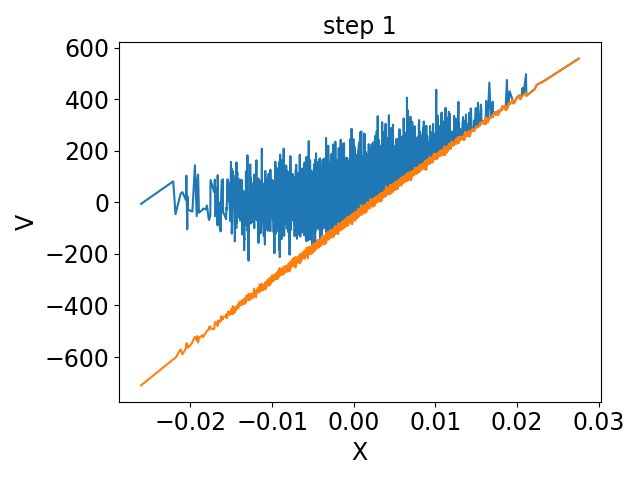}
\includegraphics[width=0.3\textwidth]{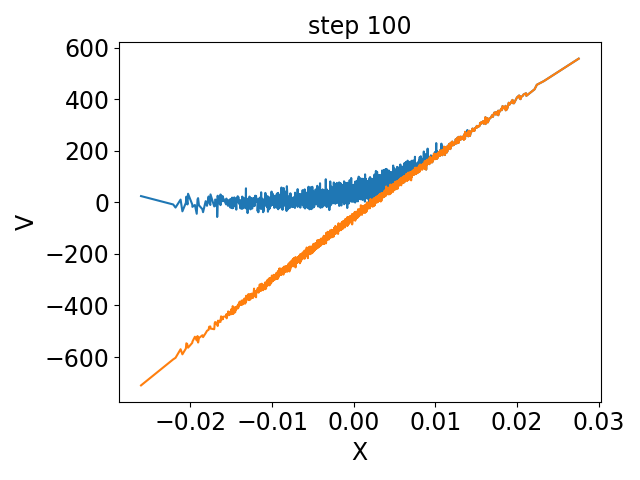}
\includegraphics[width=0.3\textwidth]{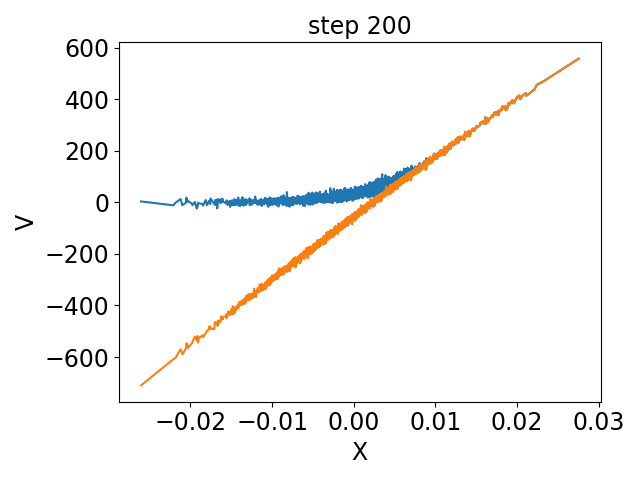}

\includegraphics[width=0.3\textwidth]{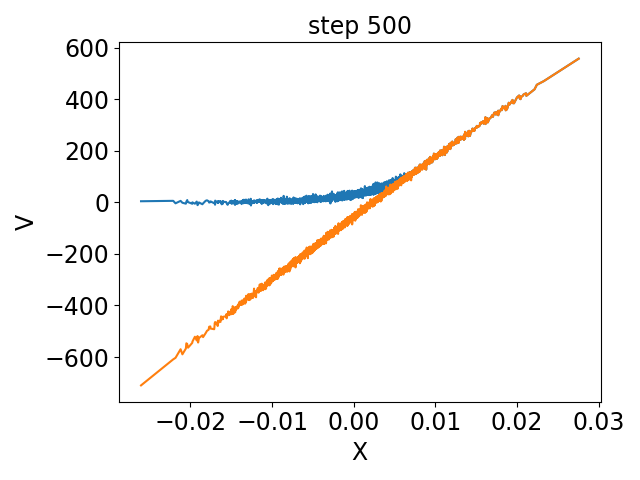}
\includegraphics[width=0.3\textwidth]{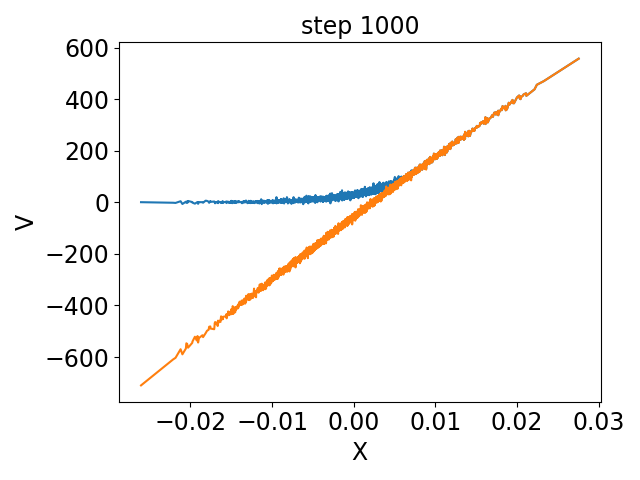}
\includegraphics[width=0.3\textwidth]{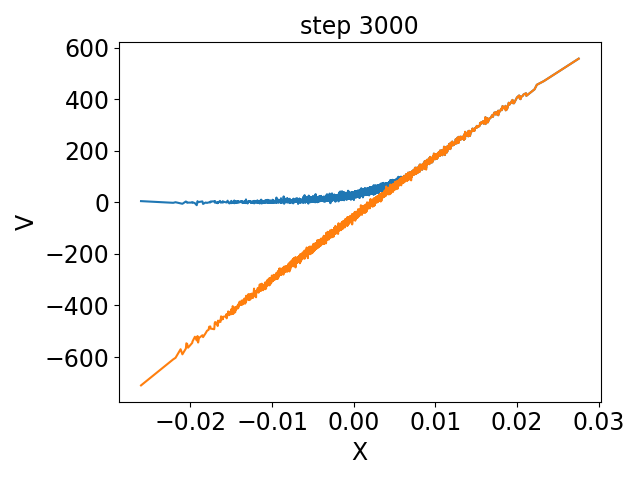}
\caption{Evolution of future values of Bermudan swaption with training (1st exercise date).  Blue lines represent the portfolio value, and orange lines the exercise value.}
\label{fig:Bermudan-training1}
\end{figure}

\begin{figure}[h]
\includegraphics[width=0.3\textwidth]{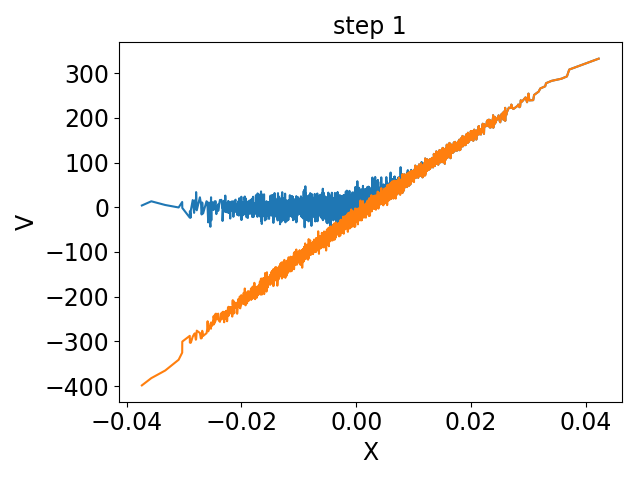}
\includegraphics[width=0.3\textwidth]{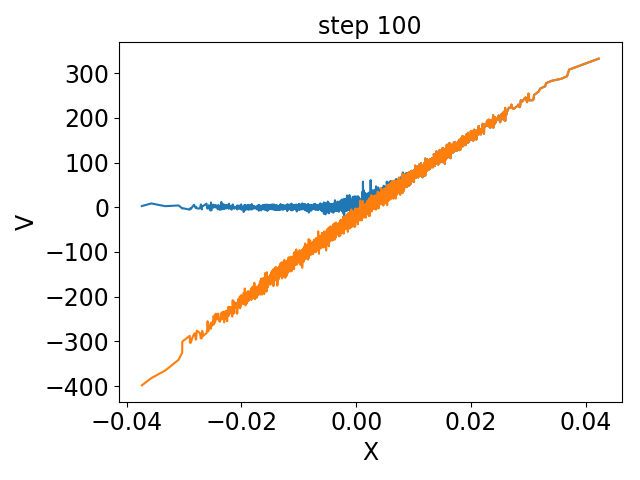}
\includegraphics[width=0.3\textwidth]{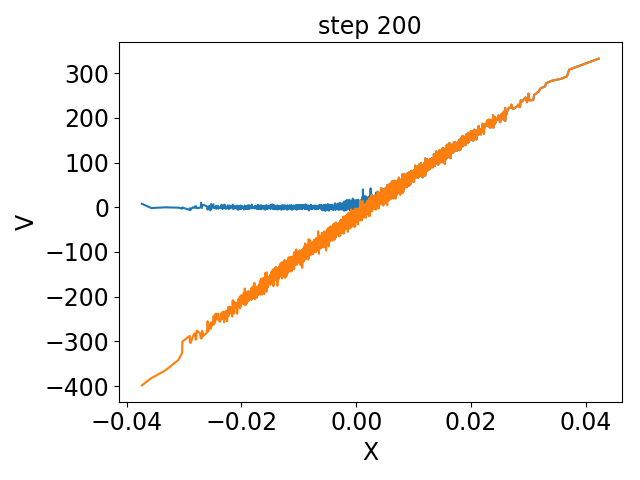}

\includegraphics[width=0.3\textwidth]{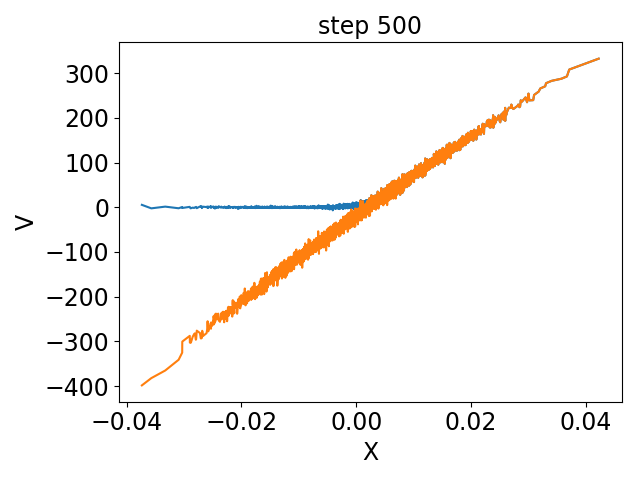}
\includegraphics[width=0.3\textwidth]{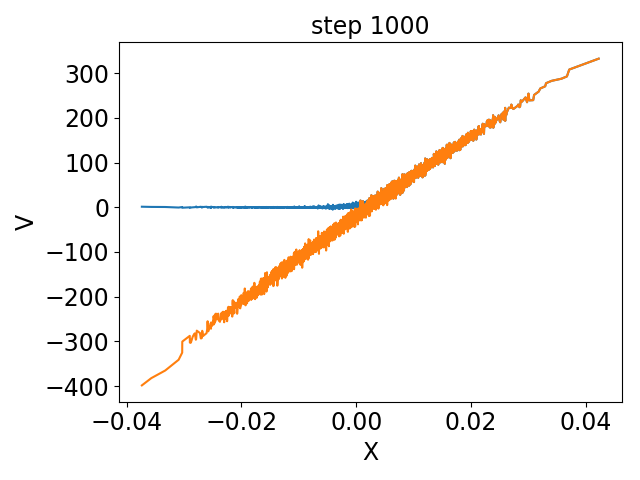}
\includegraphics[width=0.3\textwidth]{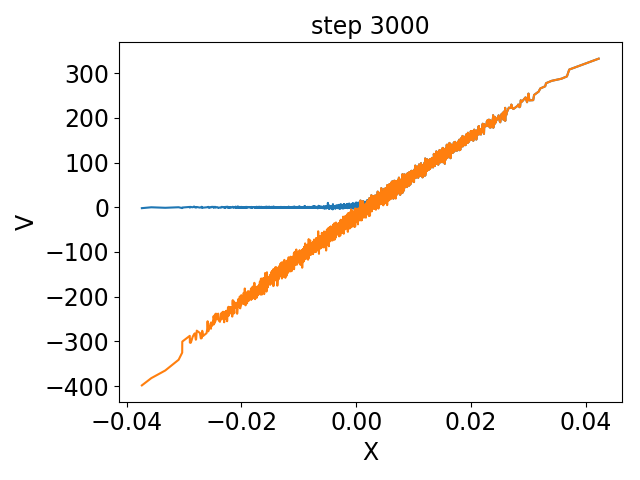}
\caption{Evolution of future values of Bermudan swaption with training (4th exercise date).}
\label{fig:Bermudan-training2}
\end{figure}

\begin{figure}[h]
\includegraphics[width=0.49\textwidth]{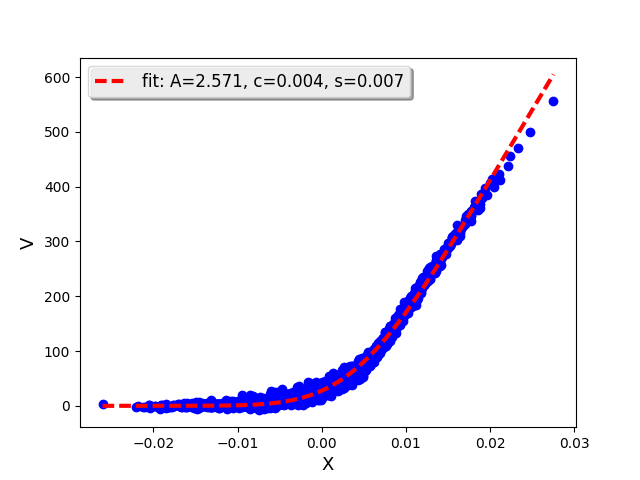}
\includegraphics[width=0.49\textwidth]{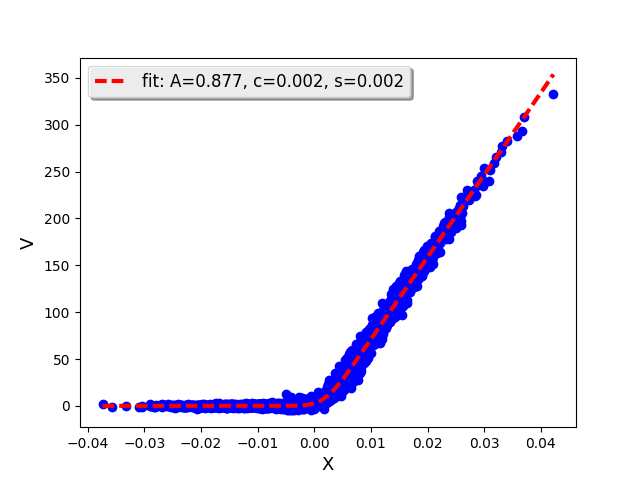}
\caption{Fitting future values of Bermudan swaptions to Bachelier formula.}
\label{fig:Bermudan-regression}
\end{figure}

\subsection{The algorithm}

As a concrete example, we consider in this section Bermudan swaption. A Bermudan swaption is an option to enter into a swap contract on a given set of dates (exercise dates). For simplicity, we consider the cash-settled case, for which the exposure vanishes after the option is exercised. The calculation of the exposure of a Bermudan swaption involves three stages. The first stage (pre-training) uses forward induction to generate all quantities that do not require knowledge of the neural network. In this stage, the following operations are carried out:
\begin{itemize}
\item Forward evolve the risk factors, and store them in a tensor $X_{ipn}$, with dimension index $i$, path index $p$, and time index $n$. 
\item Compute the exercise values, and store them in a tensor $U_{pm}$, with path index $p$, time index $m$, where $T_m$ represent the exercise dates.
\end{itemize}
The second stage is to construct and train the neural network, which uses backward induction. In this stage, the following operations are carried out:
\begin{itemize}
\item Build the neural network. At each time step $T_n$, construct a neural network for the Delta's, i.e. $\frac{\partial V}{\partial X_i}\equiv Z_i$. Each neural network is in the form of a function that maps a vector ${\mathbf X}_{pn}\equiv \left(X_{1pn}, \cdots,X_{dpn}\right)$ to a vector ${\mathbf Z}_{pn}\equiv \left(Z_{1pn},\cdots,Z_{dpn} \right)$, i.e. ${\cal F}^{(n)}:\:{\mathbf X}_{pn}\to {\mathbf Z}_{pn}$.
\item Backward induction. Evolve the future value according to the diffusion equation (\ref{eq:V-backward}) in backward form and the jump condition (\ref{eq:V-jump}) for early exercise. The results are stored in a tensor $V_{pn}$, with path index $p$, and time index $n$. 
\item Construct the loss function from $V_{pn}$. The loss function is of the form 
\begin{equation}
{\cal L}\left(V_{0},Z_i^{(0)},w_{i}^{\left(n\right)}\right)=\frac{1}{{\cal A}}\sum_{p}\left(V_{p0}-V_{0}\right)^{2}.
\end{equation}
\item Train the neural network, using for example the Adam optimizer.
\item Store the future values in the last run in a tensor $V_{pn}$. In addition, for the exposure calculation, we also need to extract the exercise indicator $\eta_{pm}$ at each exercise date $T_m$ for each path $\omega_p$, where 
\begin{equation}
\eta_{pm} = \left\{ \begin{array}{rl}
  0 &\mbox{if exercised at $T_m$,} \\
 1 &\mbox{otherwise.} 
       \end{array} \right.
\end{equation}
\end{itemize}
The third stage (post-training) is to use the stored future values and exercise indicators to compute the exposure. The following operations are carried out:
\begin{itemize}
\item Forward induction. The purpose is to obtain an indicator at each credit time $T_n$ to represent whether the option has been exercised on this path. This can be achieved by simply multiplying all the stored exercise indicators before $T_n$, i.e. $\tilde{\eta}_{pn}=\prod_{m<n}\eta_{pm}$. Here $\tilde{\eta}_{pn}=0$ if the option has been exercised before time $T_n$, and $\tilde{\eta}_{pn}=1$ otherwise.
\item Averaging. EPE/ENE at each credit date is obtained by averaging the positive/negative part of the future value for paths where the option has not been exercised:
\begin{equation}
EPE\left(T_{n}\right)=E\left[\left(\tilde{\eta}_{pn}V_{pn}\right)^{+}\right],
\end{equation}
\begin{equation}
ENE\left(T_{n}\right)=E\left[\left(\tilde{\eta}_{pn}V_{pn}\right)^{-}\right].
\end{equation}
\end{itemize}

\subsection{Hull-White one factor model}
The above algorithm is quite general, and applies to different models. Below we present a concrete example using Hull-White one factor model. The advantage of a one-factor model is that one can easily visualize the resulting functional form of the future value. We follow the notation of \cite{AndersenPiterbarg}. The short rate can be written as $r\left(t\right)=x\left(t\right)+f\left(0,t\right)$, with the initial forward rate $f\left(0,t\right)$, and the stochastic part
\begin{equation}
dx\left(t\right)=\left[y\left(t\right)-\kappa x\left(t\right)\right]dt+\sigma_r\left(t\right)dW\left(t\right),
\end{equation}
with $x(0)=0$, and 
\begin{equation}
y\left(t\right)= \int_0^t e^{-2\kappa(t-u)} \sigma_r(u)^2 du.
\end{equation}
Here the mean-reversion speed $\kappa$ is set to be a constant, while the volatility $\sigma_r(t)$ retains term structure.
 
With a single risk factor $x(t)$, the neural network is in a simple form. For example, with 2 hidden layers, it reads explicitly
\begin{equation}
 {\cal F}\left(x\right)=\sum_{j}w_{j}^C\phi_B\left(\sum_{i}w_{ji}^B\phi_{A}\left(w_{i}^A x\right)\right),
\end{equation}
at a given time $T_n$, with the parameters $w_{i}^A$, $w_{ji}^B$, $w_{j}^C$, where $i,j=1,\cdots,1+{\tilde d}$, and the nonlinear scalar functions $\phi_A,\phi_B$.

We consider a sample trade of cash-settled Bermudan swaption, which can be exercised semi-annually starting from 1.5 years to 3.5 years. The underlying is a standard fixed-for-floating swap indexed to 3M Libor rate, with notional 10000, and fixed rate 0.028. For the Hull-White model, the mean-reversion parameter is chosen to be $\kappa=0.01$, and the model is calibrated to market data on 1/18/2018. We choose a fully connected neural network with 2 hidden layers and ${\tilde d} =10$. The EPE and ENE results are shown in Figure \ref{fig:Bermudan-EPE}. ENE vanishes as the portfolio value is never negative. EPE decreases with time, displays jumps at the exercise dates, and has a convex envelope function, as the exposure on the path vanishes if the option gets exercised. We can see from the evolution of the loss function with training that the optimization procedure converges with around 500 training steps.

We further show in Figures (\ref{fig:Bermudan-training1}), and (\ref{fig:Bermudan-training2}) the evolution of the future values with training at two exercise dates. For comparison, the exercise values are displayed. The exercise value $U_n\left( x \right)$ does not depend on the neural network, and hence does not change with training. It is a linear function of $x$. The functional form of the portfolio value $V_n\left( x \right)$ evolves with training. As we started with the randomly chosen parameters $\left(V_{0},Z_i^{(0)},w_{i}^{\left(n\right)}\right)$,  $V_n\left( x \right)$ starts out quite noisy. As training proceeds, $V_n\left( x \right)$ converges to a smooth function at about 500 training steps. The resulting function increases monotonically with $x$, and interpolates between the $x$-axis and the exercise value. Such a functional form is typical for an option price: when the interest rate is high, the underlying swap is deep in the money, the swaption will be exercised, and the swaption value will approach the exercise value; when the interest rate is low, the swap is deep out of the money, and the swaption will not be exercised, and the swaption value will be close to zero.

We can fit the resulting function to a Bachelier-type option price formula:
\begin{equation}
V_{Bach}\left(x\right)=A\left[\left(x-c\right)\Phi\left(\frac{x-c}{s}\right)+s\phi\left(\frac{x-c}{s}\right)\right],
\end{equation}
where $\Phi\left(\cdot\right)$  is the standard Gaussian cumulative distribution function, and $\phi\left(\cdot\right)$ the corresponding probability density function. As $x\to\infty$, $V_{Bach}\left(x\right)\to A\left(x-c\right)$. So $A$ basically corresponds to the slope of the exercise value, and $c$ its intersection with $x$-axis. $s$ represents an effective volatility, and larger $s$ gives more rounded interpolation between the $x$-axis and the exercise value. The results are shown in Figure \ref{fig:Bermudan-regression}.

Being able to visualize the functional form of the future value can shed new light on other approaches. For AMC, a difficulty is the choice of basis functions. As the future value of Bermudan swaption is of the form of an option price, while the exercise value is essentially linear, merely using powers of the exercise value as basis functions seems not an optimal choice, and related European option prices should also be included. Furthermore, with the known functional form of the future value, one can try to replace the linear regression step in AMC by directly fitting the inferred function, e.g. Bachelier-type formula for Bermudan swaption. Note that it is important to constrain the parameters to be in the financially meaningful regimes when carrying out the fitting, as the result is a local minimum, and not a global minimum.

\section{MtM cross-currency swap}

\begin{figure}[h]
\includegraphics[width=0.49\textwidth]{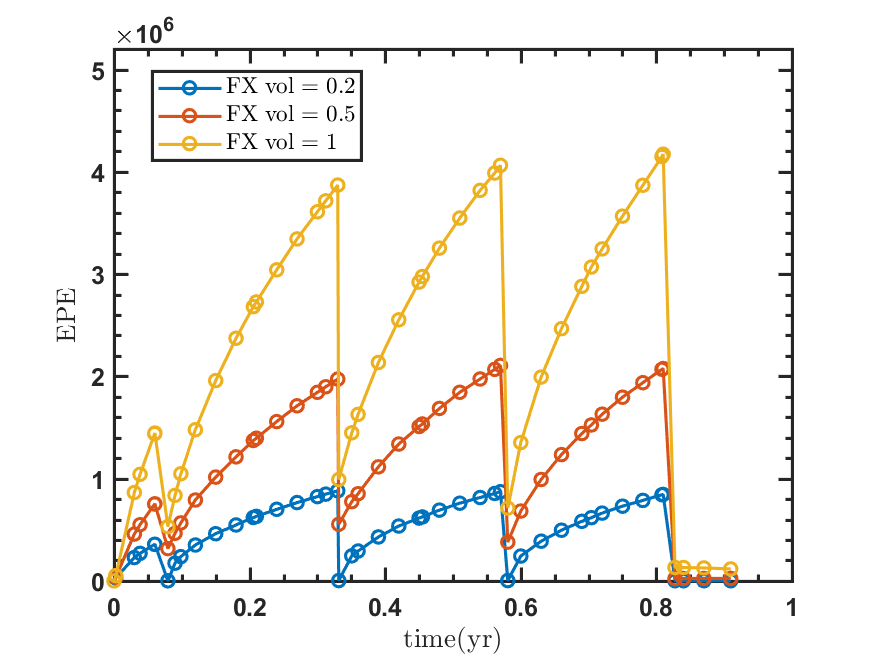}
\includegraphics[width=0.49\textwidth]{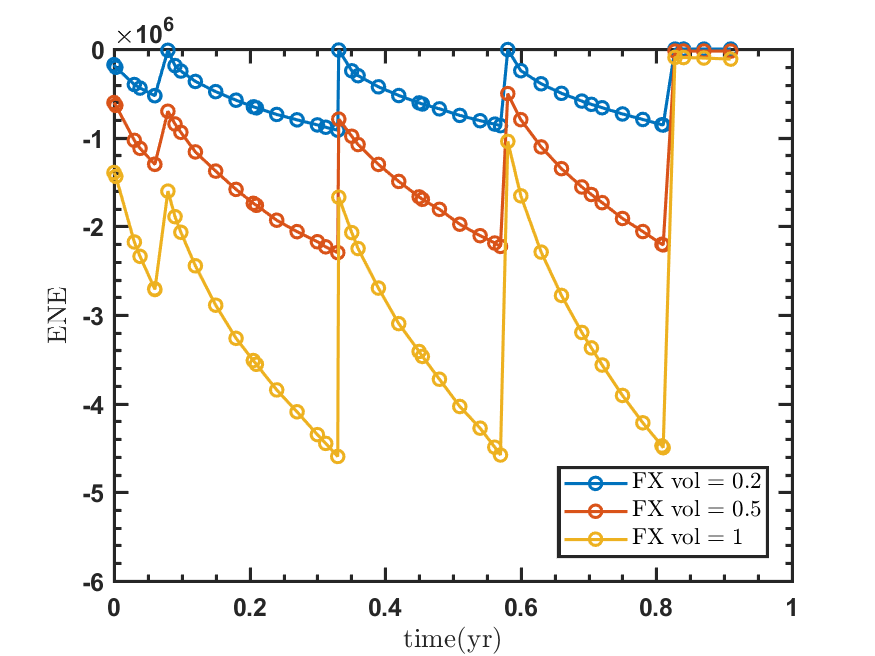}
\caption{Exposure of MtM XCCY swap with $\left(\sigma_{0},\sigma_{1},\eta_{1}\right)=\left(0.001,0.001,0.2\right),\left(0.1,0.1,0.5\right),\left(0.2,0.2,1\right)$.}
\label{fig:Exposure-of-MtM-un-stress-NN2}
\end{figure}

\begin{figure}[h]
\includegraphics[width=0.49\textwidth]{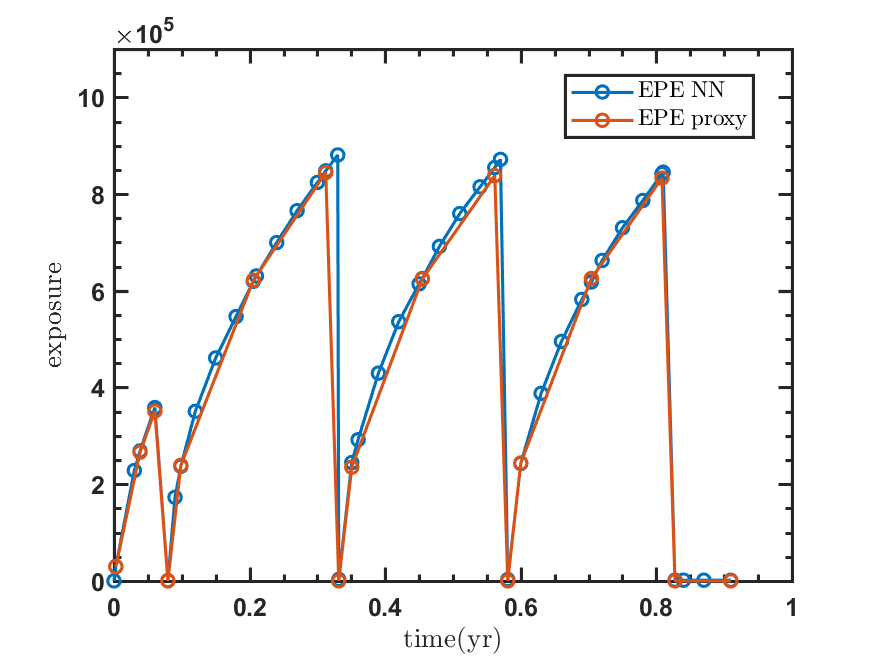}
\includegraphics[width=0.49\textwidth]{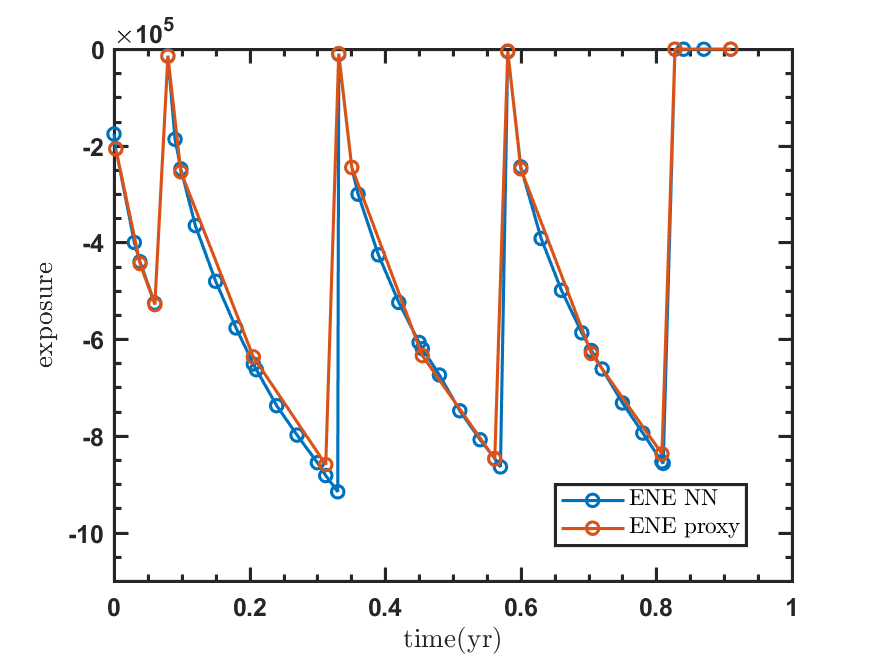}

\includegraphics[width=0.49\textwidth]{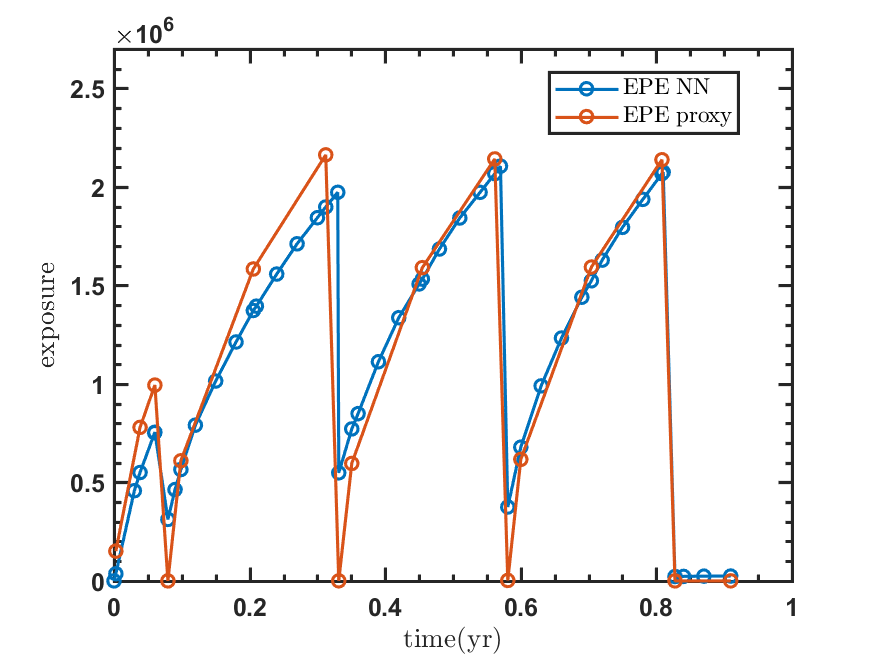}
\includegraphics[width=0.49\textwidth]{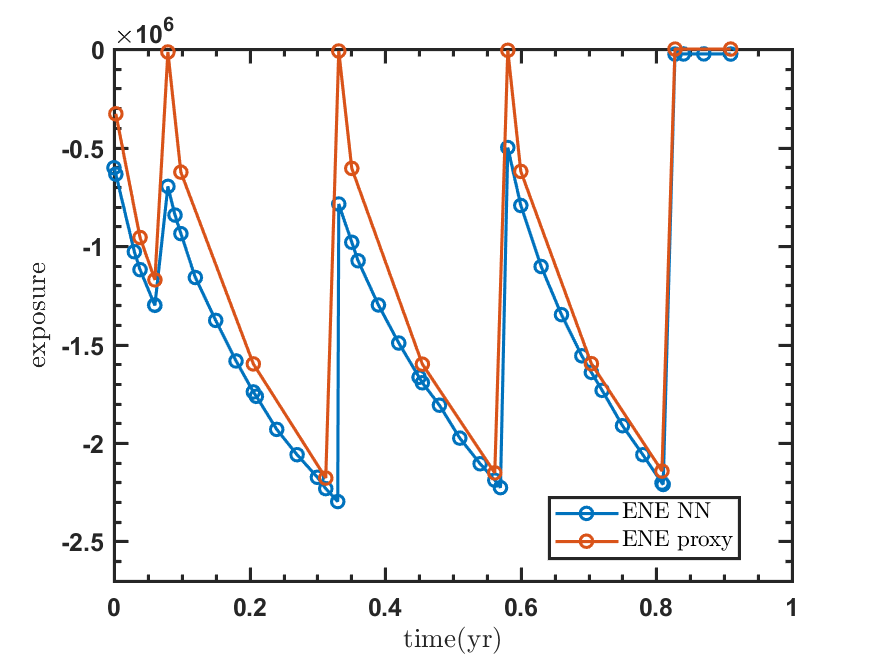}
\caption{Comparison of exposure from NN method and proxy method for MtM XCCY swap for different volatilities. Up:  $\left(\sigma_{0},\sigma_{1},\eta_{1}\right)=\left(0.001,0.001,0.2\right)$. Down: $\left(\sigma_{0},\sigma_{1},\eta_{1}\right)=\left(0.2,0.2,1\right)$.}
\label{fig:Exposure-of-MtM-un-stress-NN0}
\end{figure}

\begin{figure}[h]
\includegraphics[width=0.3\textwidth]{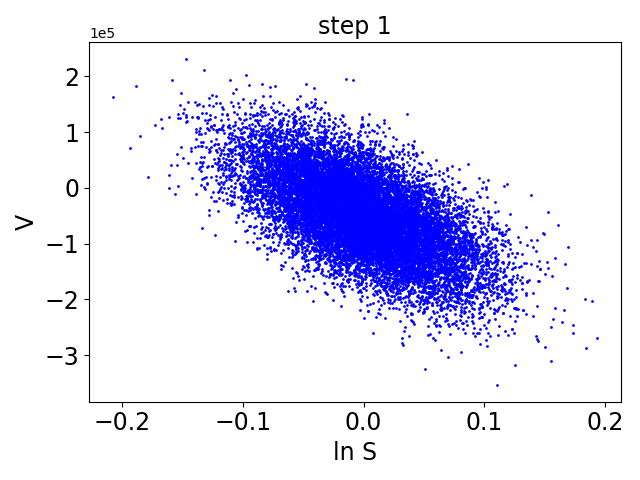}
\includegraphics[width=0.3\textwidth]{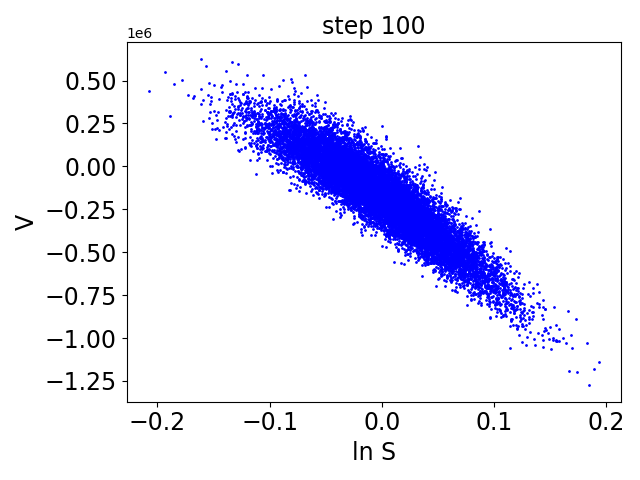}
\includegraphics[width=0.3\textwidth]{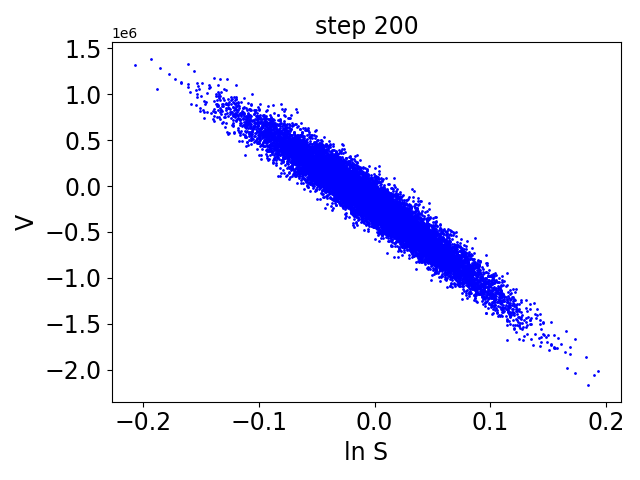}

\includegraphics[width=0.3\textwidth]{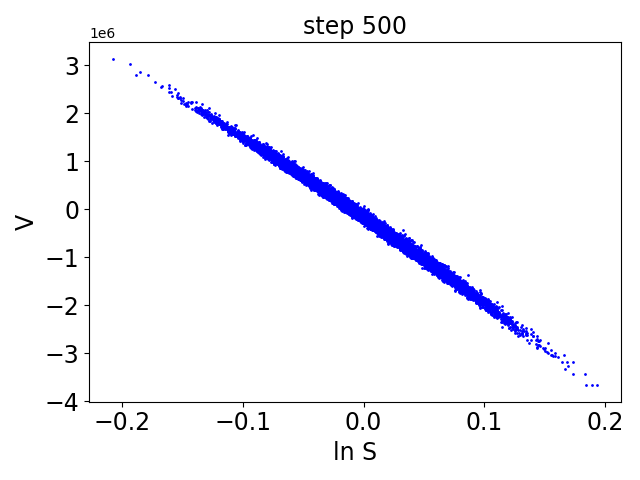}
\includegraphics[width=0.3\textwidth]{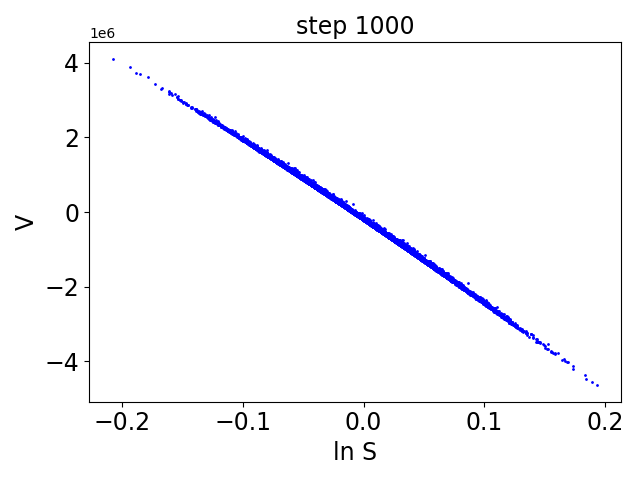}
\includegraphics[width=0.3\textwidth]{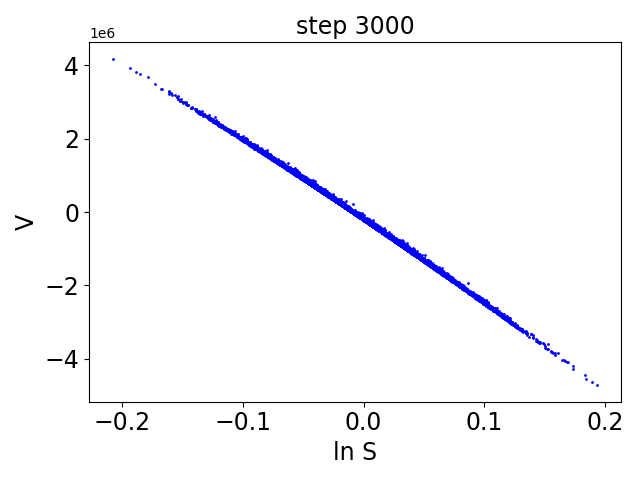}
\caption{Evolution of future values of MtM XCCY swap with training. Here $\left(\sigma_{0},\sigma_{1},\eta_{1}\right)=\left(0.001,0.001,0.2\right)$. }
\label{fig:XCCY-training1}
\end{figure}

\begin{figure}[h]
\includegraphics[width=0.3\textwidth]{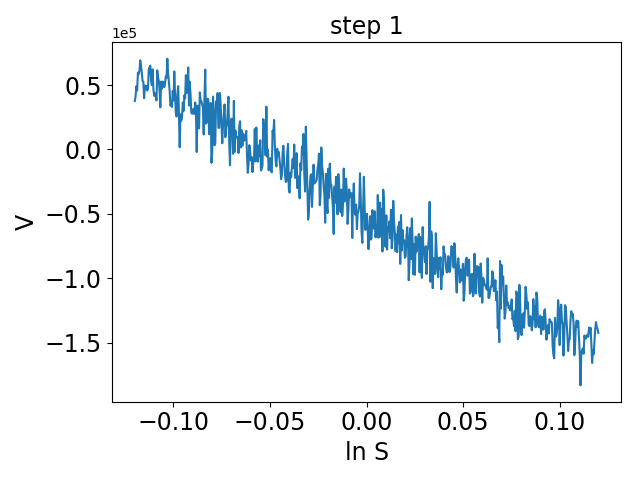}
\includegraphics[width=0.3\textwidth]{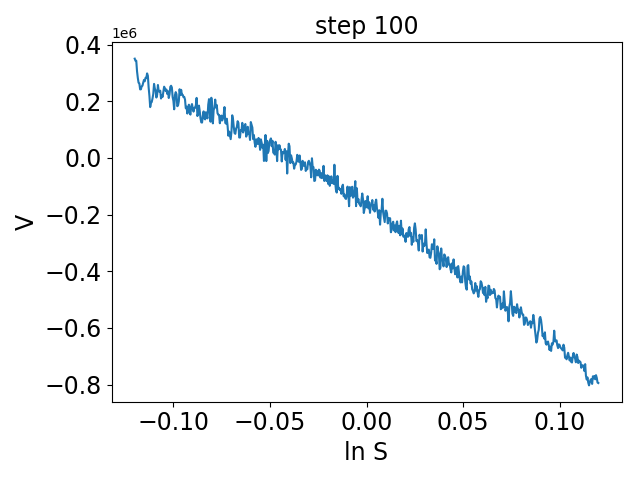}
\includegraphics[width=0.3\textwidth]{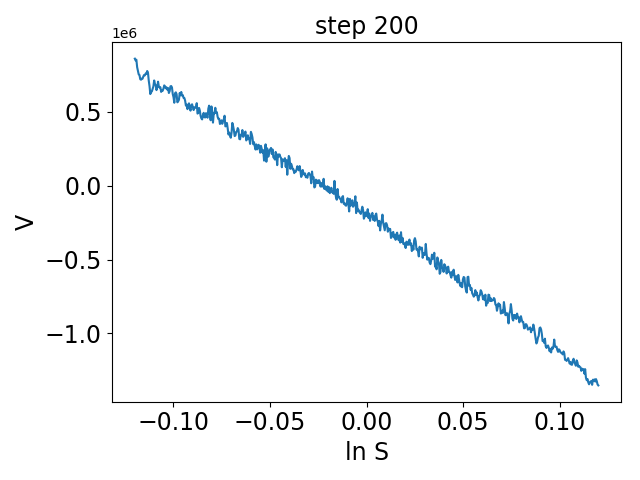}

\includegraphics[width=0.3\textwidth]{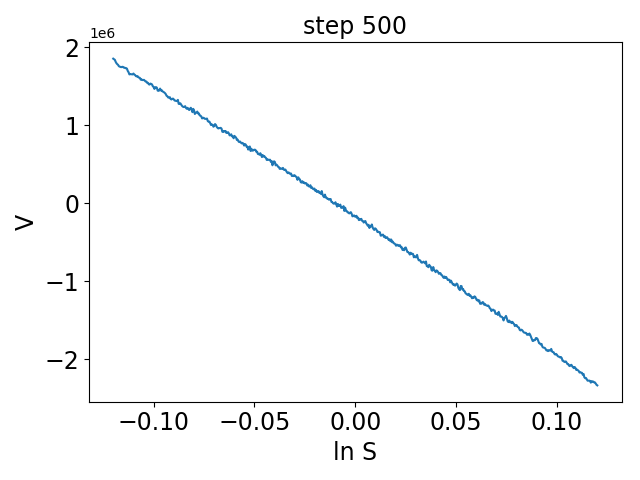}
\includegraphics[width=0.3\textwidth]{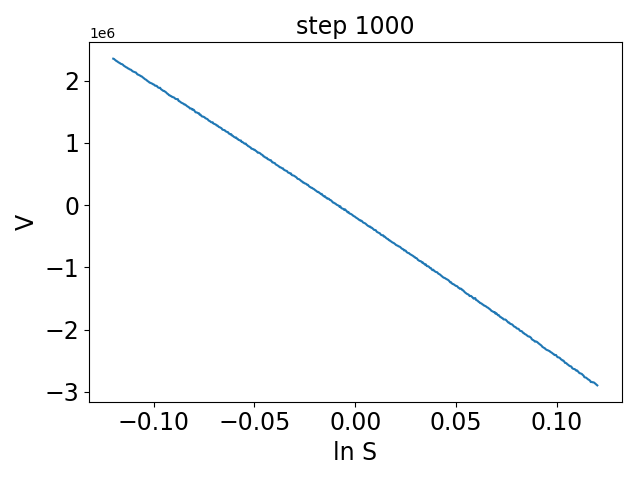}
\includegraphics[width=0.3\textwidth]{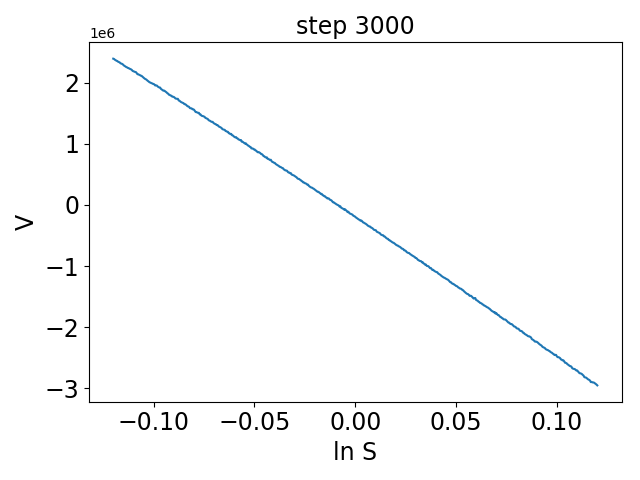}
\caption{Evolution of future values of MtM XCCY swap with training (projected to $x_{0}=0$, $x_{1}=0$). Here $\left(\sigma_{0},\sigma_{1},\eta_{1}\right)=\left(0.001,0.001,0.2\right)$. }
\label{fig:XCCY-training11}
\end{figure}

\begin{figure}[h]
\includegraphics[width=0.3\textwidth]{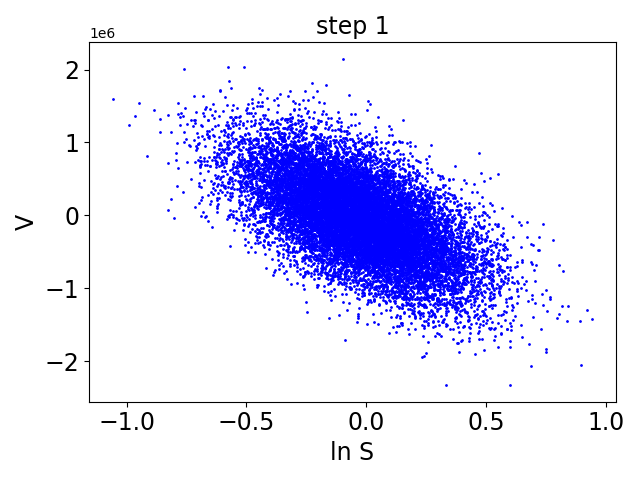}
\includegraphics[width=0.3\textwidth]{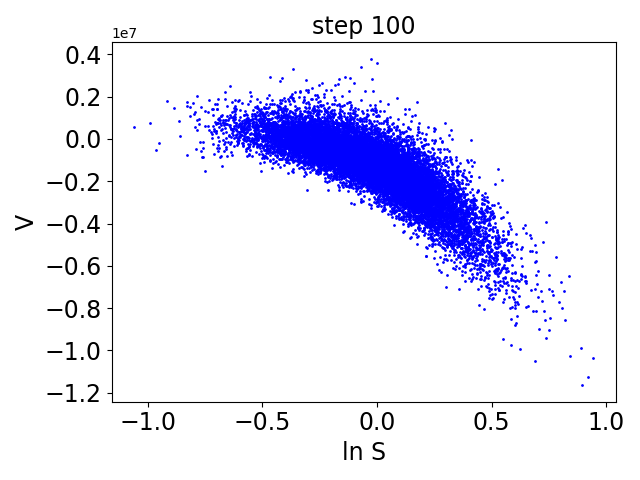}
\includegraphics[width=0.3\textwidth]{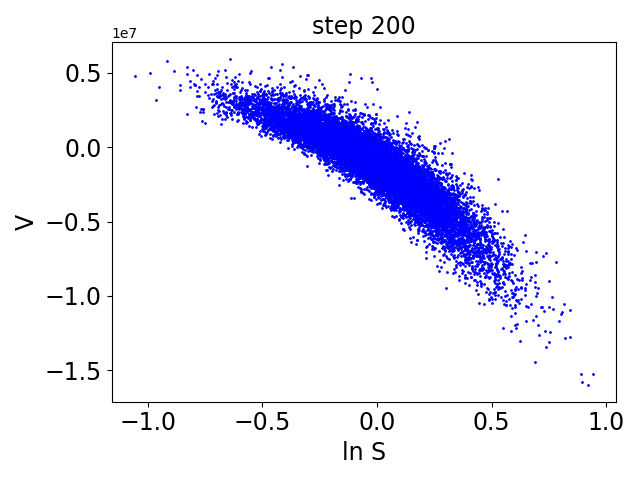}

\includegraphics[width=0.3\textwidth]{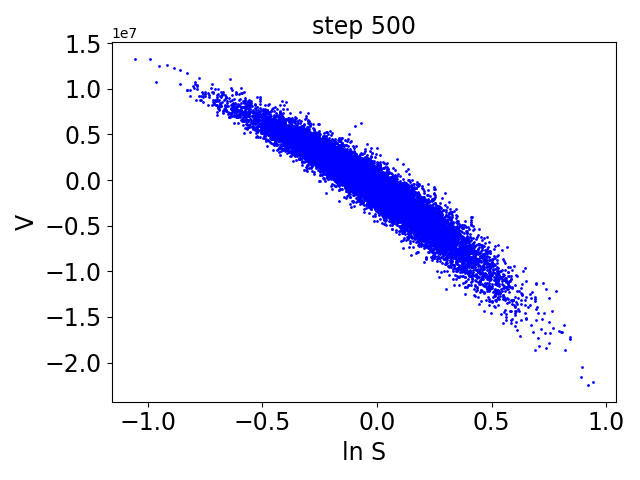}
\includegraphics[width=0.3\textwidth]{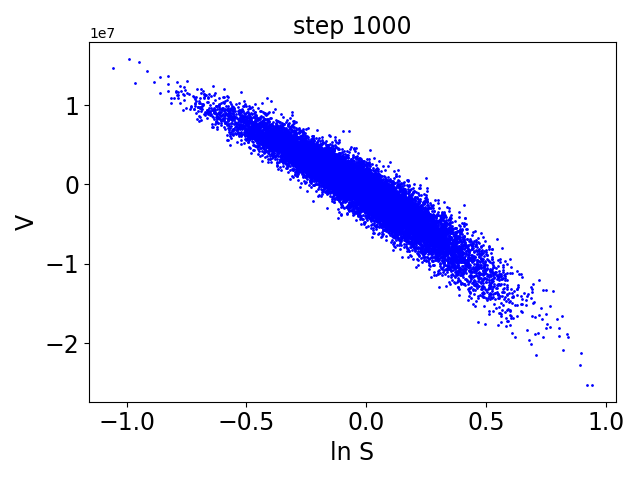}
\includegraphics[width=0.3\textwidth]{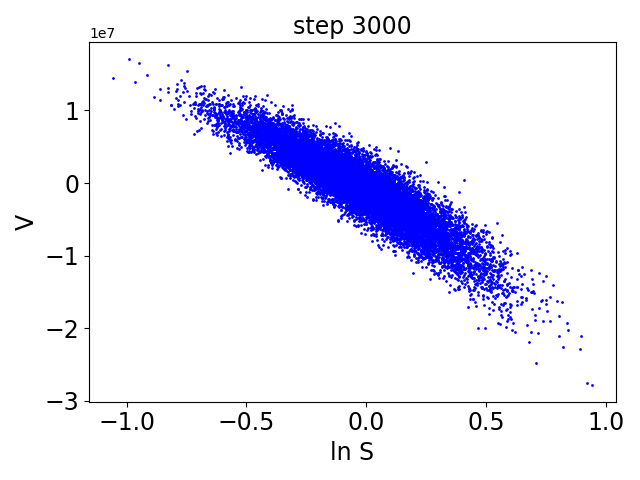}
\caption{Evolution of future values of MtM XCCY swap with training. Here $\left(\sigma_{0},\sigma_{1},\eta_{1}\right)=\left(0.2,0.2,1\right)$. }
\label{fig:XCCY-training2}
\end{figure}

\begin{figure}[h]
\includegraphics[width=0.3\textwidth]{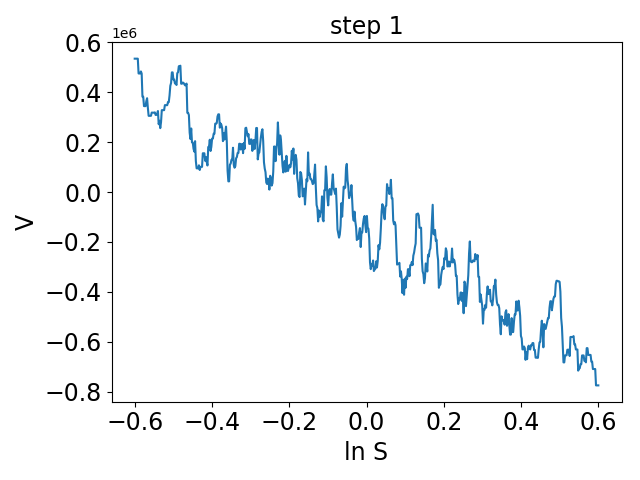}
\includegraphics[width=0.3\textwidth]{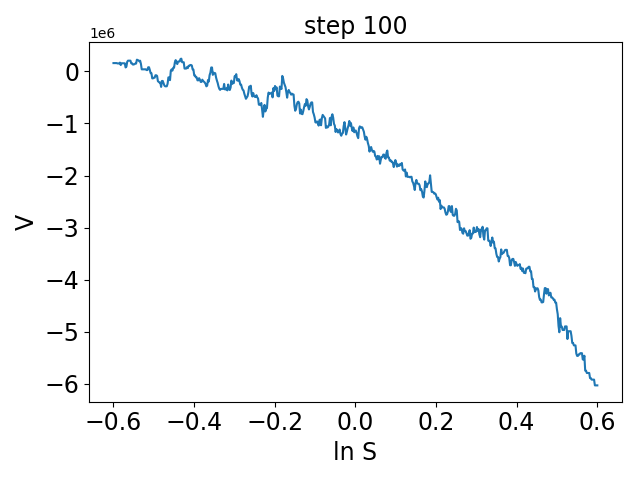}
\includegraphics[width=0.3\textwidth]{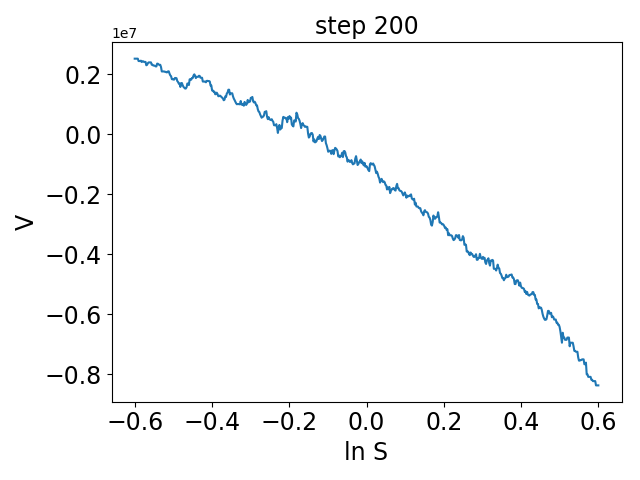}

\includegraphics[width=0.3\textwidth]{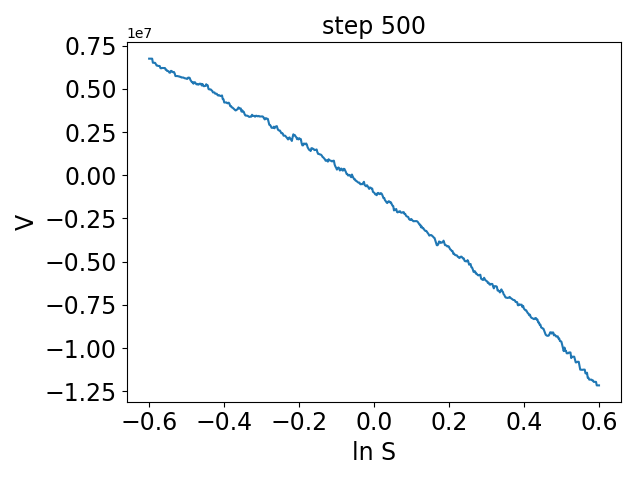}
\includegraphics[width=0.3\textwidth]{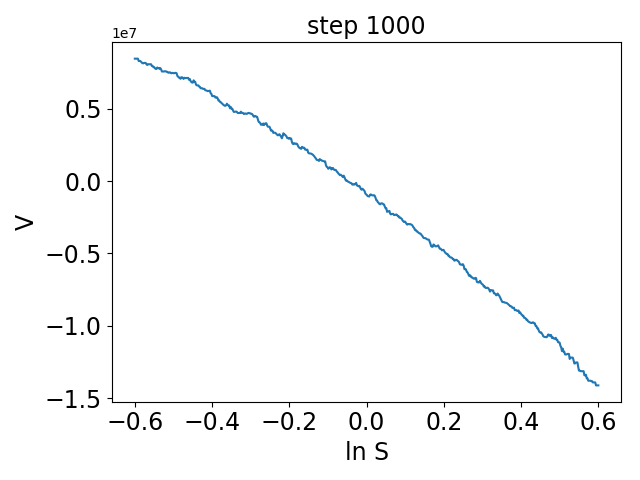}
\includegraphics[width=0.3\textwidth]{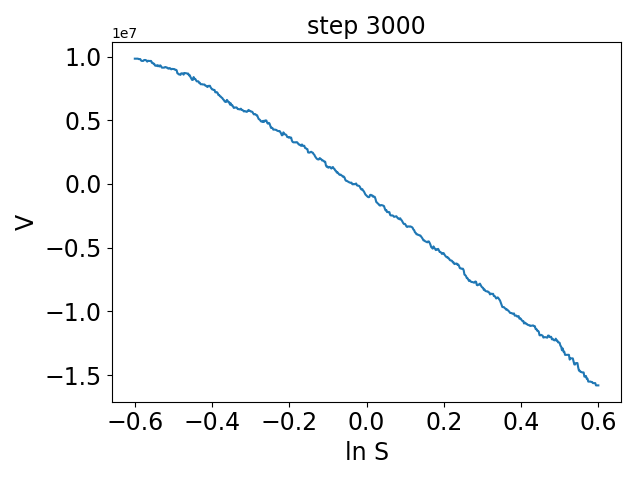}
\caption{Evolution of future values of MtM XCCY swap with training (projected to $x_{0}=0$, $x_{1}=0$). Here $\left(\sigma_{0},\sigma_{1},\eta_{1}\right)=\left(0.2,0.2,1\right)$. }
\label{fig:XCCY-training22}
\end{figure}

\begin{figure}[h]
\includegraphics[width=0.49\textwidth]{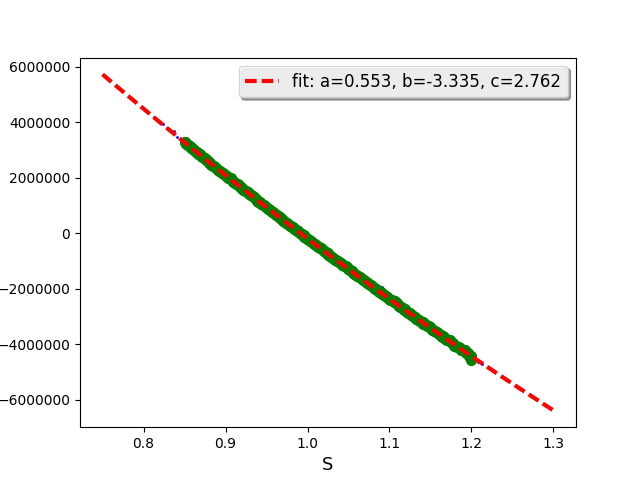}
\includegraphics[width=0.49\textwidth]{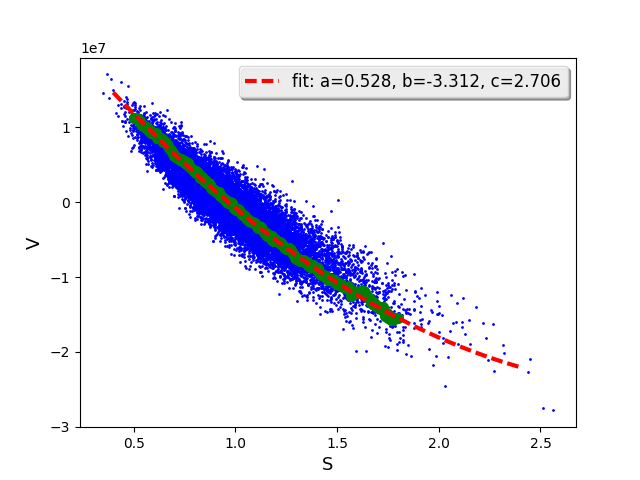}
\caption{Regression of future value for MtM XCCY swap, fitting to function $V=10^{7}\left(aS^{2}+bS+c\right)$. Blue dots are raw data of $\left(S,V\right)$ with random $x_0, x_1$, green lines are $\left(S,V\right)$ projected to $x_{0}=0, x_{1}=0$, red lines are fitted function of the projected data.}
\label{fig:XCCY-fit}
\end{figure}

The advantage of the neural network approach lies in its easiness to generalize to high dimensions, i.e. multi-factor models. In this section, we consider as example a Mark-to-Market (MtM) cross-currency (XCCY) swap. The exposure of cross-currency swap involves risk factors from multiple asset classes, i.e. interest rate (IR) and foreign exchange (FX), hence it serves as a prototype for cross-asset CVA modeling. In addition, MtM swaps reset the notional periodically. Consequently in general their future values do not have analytic expressions, and they serve as testing ground for the neural network method.

\subsection{The model}
We consider the correlated dynamics of both IR and FX rates. The IR short rates are again decomposed into the initial curve part and the stochastic part: $r_{i}\left(t\right)=x_{i}\left(t\right)+f_{i}\left(0,t\right)$. Under the domestic risk-neutral measure, the risk factor evolution is as follows:
\begin{eqnarray}
dx_{0}\left(t\right)&=&\left[y_{0}\left(t\right)-\kappa_{0}x_{0}\left(t\right)\right]dt+\sigma_{0}\left(t\right)dW_{0}\left(t\right), \\
dx_{i}\left(t\right)&=&\left[y_{i}\left(t\right)-\kappa_{i}x_{i}\left(t\right)-\rho_{i,i+N}\sigma_{i}\left(t\right)\eta_{i}\left(t\right)\right]dt+\sigma_{i}\left(t\right)dW_{i}\left(t\right), \\
d\ln S\left(t\right)&=&\left[r_{0}\left(t\right)-r_{i}\left(t\right)-\frac{1}{2}\eta_{i}\left(t\right)^{2}\right]dt+\eta_{i}\left(t\right)dW_{i+N}\left(t\right),
\end{eqnarray}
with $<dW_{i}\left(t\right),dW_{j}\left(t\right)>=\rho_{ij}dt$, where $i,j=0,\cdots,2N$, and $N$ is the number of foreign currencies. Note that the $\rho\sigma\eta$ cross term comes from change of measure from the foreign risk-neutral measure to the domestic risk-neutral measure. In the domestic risk-neutral measure, the numeraire is $B\left(t\right)=\exp\left[\int_{0}^{t}r_{0}\left(u\right)du\right]$.

To set the stage, we consider first a standard XCCY swap with one floating leg in domestic currency and one fixed leg in foreign currency. Standard XCCY swaps are linear products, and their future values can be determined analytically from the information available at credit time $t$. The future value of the fixed leg is 
\begin{equation}
V_{fxd}\left(t\right)=N_f S\left(t\right)\sum_{n} K_{n}\tau_{n}P_{D}\left(t,T_n^{P}\right),
\end{equation}
with notional $N_f$, fixed coupon rate $K_n$, day counting for each accrual period $\tau_n$, the foreign exchange rate $S\left(t\right)$, and the discount factor $P_{D}\left(t,T_n^{P}\right)$ from $t$ to future payment time $T_n^{P}$. The future value of the  floating leg is 
\begin{equation}
V_{flt}\left(t\right)=N_f S\left( 0\right)\sum_{n}\left[\alpha_{n}L_{n}\left(t\right)+\beta_{n}\right]\tau_{n}P_{D}\left(t,T_{n}^{P}\right),
\end{equation}
with multiplier $\alpha_{n}$, spread $\beta_{n}$, and the forward rate $L_{n}\left(t\right)$. There are typically notional exchanges at maturity, which can be included by replacing $K_n\tau_n\to 1+K_n\tau_n$, and $\beta_n\tau_n\to 1+\beta_n\tau_n$ for the last period.

The story is quite different for cross-currency swaps with MtM legs. MtM XCCY swaps reset the notional at the beginning of each accural period. The MtM leg includes two payments: (1) notional payment: $N_{f}\left[S\left(T_{n}\right)-S\left(T_{n+1}\right)\right]$, (2) rate payment: $N_{f}S\left(T_{n}\right)L_{n}\tau_{n}$, which in addition to having a floating rate $L_{n}$, effectively also has a stochastic notional $N_{f}S\left(T_{n}\right)$. Hence the future value of the MtM leg is
\begin{equation}
V_{mtm}\left(t\right)=E_{t}^{Q}\left[N_{f}\sum_{n}P_{D}\left(t,T_{n}^{P}\right)\left[S\left(T_{n}\right)\left(1+L_{n}\tau_{n}\right)-S\left(T_{n+1}\right)\right]\right].
\end{equation}
One can see that the future value of MtM leg depends on the FX rates at all future reset times, i.e. $S\left(T_{n}\right)$, which are unknown at time $t$. No analytical expression for the future value of a MtM XCCY swap with correlated factors is currently known. 

One can consider a proxy model. Assuming the decoupling of the IR part and the FX part, the expectation value of the FX rates then can be computed directly using the relation 
\begin{equation}
E_{t}^{Q}\left[S\left(T_{n}\right)\right]=S\left(t\right)\frac{P_{f}\left(t,T_{n}\right)}{P_{d}\left(t,T_{n}\right)} ,
\end{equation}
 with the domestic and foreign discount factors $P_{d}\left(t,T_{n}\right) , P_{f}\left(t,T_{n}\right)$. Such a proxy method produces an analytic expression for the future value, but fails to capture the convexity adjustment due to the cross-asset correlation.

\subsection{Neural network approach}
We will attack this problem in the NN approach. Let us start by formulating the problem in the BSDE framework. Consider a XCCY swap with one MtM leg and one floating leg. The future value is $V_{XCCY}\left(t\right)=V_{mtm}\left(t\right)-V_{flt}\left(t\right)$. In the domestic risk-neutral measure, it grows with rate $r_{0}\left(t\right)$, and hence follows the BSDE:
\begin{equation}
dV_{\rm XCCY}\left(t\right)=r_{0}\left(t\right)V_{\rm XCCY}\left(t\right)dt+\sum_{i}\frac{\partial V_{\rm XCCY}}{\partial X_{i}}\left(t,\mathbf{X}_{t}\right)\sigma_{i}\left(t\right)dW_{i}\left(t\right).
\label{eq:V-XCCY}
\end{equation}
Here we group the risk factors in a vector $\mathbf{X}\equiv \left(x_0, x_1, \ln S  \right)$, and the corresponding volatilities in a vector $\mathbf{\sigma}\equiv\left(\sigma_0, \sigma_1, \eta_1 \right)$.
We also include the jump condition at the cashflow dates
\begin{equation}
V_{\rm XCCY}\left(T_{n}^{+}\right)=V_{\rm XCCY}\left(T_{n}^{-}\right)-CF\left(T_{n}\right),
\label{eq:CF-jump}
\end{equation}
and the boundary condition at maturity 
\begin{equation}
V_{\rm XCCY}\left(T_{N}^{+}\right)=0.
\end{equation}

We observe that while the cashflow $CF\left(T_{n}\right)$ is fully determined by information available at time $T_{n}$, the Delta's $\frac{\partial V_{\rm XCCY}}{\partial X_{i}}\left(t,\mathbf{X}_{t}\right)$ depend on information beyond time t, in particular the future FX rates. We proceed by parameterizing the Delta's of the future value by fully connected neural networks: 
\begin{equation}
\frac{\partial V_{\rm XCCY}}{\partial X_{i}}\left(T_{n},\mathbf{X}_n \right)\simeq{\cal F}_{i}^{\left(n\right)}\left(X_{1},\cdots,X_{d}\right).
\end{equation}
Below we present the algorithm for the exposure calculation in the NN approach. 

First stage (pre-training): 
\begin{itemize}
\item Forward evolve the risk factors, and store them in a tensor $X_{ipn}$, with dimension index $i$, path index $p$, and time index $n$. 
\end{itemize}

Second stage:
\begin{itemize}
\item Build the neural network. At each time step, construct a neural network for the Delta's, i.e. $\frac{\partial V_{\rm XCCY}}{\partial X_i}\equiv Z_i$. 
\item Forward induction. Evolve the future value according to the diffusion equation (\ref{eq:V-XCCY}) in forward form and the jump condition (\ref{eq:CF-jump}) for cashflow. The results are stored in a tensor $V_{pn}$, with path index $p$, and time index $n$. 
\item  Construct the loss function from $V_{pn}$. The loss function is of the form 
\begin{equation}
{\cal L}\left(V_{0},Z_i^{(0)},w_{i}^{\left(n\right)}\right)=\frac{1}{{\cal A}}\sum_{p}V\left(T_{N}^+,\omega_{p}\right)^{2}. 
\end{equation}
\item Train the neural network.
\end{itemize}

Third stage (post-training): 
\begin{itemize}
\item Averaging.  Compute EPE/ENE at each credit date by averaging the positive/negative part of the future values.
\end{itemize}

Since no early exercise is involved in the current problem, we use forward induction to evolve the future value. The exposure calculation is also more straightforward.

We consider a MtM XCCY swap with currency pair CAD/USD, maturity 0.83 year, quarterly paid and reset. The initial FX rate is 0.76. The Hull-White model has mean-reversion $\kappa=0.01$ for both currencies, and flat yield curve $f_{\rm USD}\left(0,t\right)=0.01$, $f_{\rm CAD}\left(0,t\right)=0.02$. The correlation parameters are: $\rho_{{\rm USD}, {\rm CAD}}=0.149$, $\rho_{{\rm USD}, {\rm CAD/USD}}=0.139$, $\rho_{{\rm CAD}, {\rm CAD/USD}}=0.676$.  We choose a fully connected neural network with 2 hidden layers of dimension $d+{\tilde d}$, where $d=3$ (number of risk factors), and ${\tilde d} =10$. 

The resulting EPE and ENE for several different parameter sets of volatilities are shown in Figure \ref{fig:Exposure-of-MtM-un-stress-NN2}. One can see that when the volatilities are small, the exposure vanishes at the MtM dates. The practice of Mark-to-Market reduces the long-term risk to short term (here three months). As the volatilities increase, the exposure on those dates increases to finite values.  We further compare in Figure \ref{fig:Exposure-of-MtM-un-stress-NN0} the results from NN approach with those from the proxy method where cross-asset correlation is ignored. We note that proxy method always gives vanishing exposure at the MtM dates, and the NN approach properly captures the convexity adjustment missing in the proxy method. While such effects are mild in normal conditions, they can become significant in stress conditions.

We then study the functional form of future values. To have an intuitive understanding of the functional form, we try to visualize it. While it is easy to visualize the function for a single risk factor, it is more involved for multiple risk factors. As FX rate is the dominant risk factor, we study the functional relation between future value and FX rate. We present two types of plots here. The first type plots $\left(\ln S, V\right)$ for random $\left(x_{0},x_{1}\right)$ as obtained from the simulation paths. The second type plots $\left(\ln S, V\right)$ projected to given $x_{0}$, $x_{1}$. The projection is carried out using k-nearest neighbor regression, which is a non-parametric method without assuming any particular relationship among the different variables. 

The results for two sets of volatility parameters are shown in Figures \ref{fig:XCCY-training1},  \ref{fig:XCCY-training11}, \ref{fig:XCCY-training2} and \ref{fig:XCCY-training22}. When the IR volatilies are small (here $\sigma_{0} =\sigma_{1} =0.001$, and $\eta_1=0.2$), the dynamics is dominated by the FX part, and the problem is essentially one dimensional. Consequently $\left(\ln S, V\right)$ pairs converge to one-dimensional lines in both plots.  When both IR and FX volatilities are large (here $\sigma_{0} =\sigma_{1} =0.2$, and $\eta_1=1$), the problem is intrinsically high dimensional. In the random-$x$ plot, $\left(\ln S, V\right)$ pairs still form two dimensional areas after training. Such broadening effect (analogous to self-energy effect in physics) signals the important role played by IR rates in determining the functional form of future value. In the projected-$x$ plot, $\left(\ln S, V\right)$ pairs converge to one-dimensional curves. These projected curves give a glimpse of the high dimensional function $V_{\rm XCCY}\left(x_0, x_1,S \right)$.

We examine further the resulting functional dependence of future value on the FX rate $S$. In the proxy method, future value is linear in $S$. For the NN approach, to see the effect of convexity adjustment, we fit the future value to a quadratic function of $S$ (see Figure \ref{fig:XCCY-fit}). We consider $\left(S,V\right)$ pairs projected to given $x_0, x_1$. From the fitted coefficients (in particular the ratio $a/b$), one can see that convexity effect clearly exists for both small and large volatilities, though the effect is less visible for small volatilities.

\section{Conclusions}
We have explored a new approach to model the future value and compute CVA/DVA, employing neural network as a universal approximator. The core idea is common in artificial intelligence: to convert a complicated problem to a search problem.  For the present approach, it can be summarized as parameterize and optimize. The gradients of the future values are parameterized by neural network, and then efficient optimization algorithms are used to determine the involved parameters.  Immediate future directions include (1) generalizing the models to cover more risk factors, e.g. using Libor-market model (\cite{Qi18}), (2) exploring different ways of parameterization, e.g. directly parameterizing the future values (\cite{Raissi18}), (3) exploring different types of neural networks, e.g. convolutional neural network (CNN \cite{LeCun89}), generative adversarial network (GAN \cite{Goodfellow14}), (4) using this approach to model future greeks and compute MVA.   

The idea of randomization is crucial here. In some sense, the present approach is a natural generalization of American Monte Carlo. As the NN approach is more general, and does not assume prior knowledge as in AMC, it can not compete with AMC in speed. We regard this approach as first a benchmark model for AMC. What is more interesting is to combine the two approaches to form a new strategy for exotics/XVA modeling, which takes two steps: (1) use the NN method to infer the functional form of future values for each class of products, (2) then use the learned knowledge to compute the future values in a faster method like AMC or its variant. The NN step is run infrequently, and hence requirement on its speed is not stringent. The AMC step is spared of expert input. We regard this approach as a more practical way of applying deeping learning to security pricing.

\newpage{}


\begin{thebibliography}{10}
\bibitem[1]{Longstaff01}Francis A. Longstaff, and Eduardo S. Schwartz.
\textit{Valuing American Options by Simulation: A Simple Least-Squares Approach},
The Review of Financial Studies, Volume 14, Issue 1, 1 January 2001,
Pages 113\textendash 147.

\bibitem[2]{Weinan17}Weinan E, Jiequn Han, and Arnulf Jentzen. \textit{Deep
learning-based numerical methods for high-dimensional parabolic partial
differential equations and backward stochastic differential equations},
A. Commun. Math. Stat. (2017) 5: 349.

\bibitem[3]{Jiequn17}Jiequn Han, Arnulf Jentzen, and Weinan E. \textit{Overcoming
the curse of dimensionality: Solving high-dimensional partial differential
equations using deep learning}, arXiv:1707.02568.

\bibitem[4]{Goodfellow16} Ian Goodfellow, Yoshua Bengio, Aaron Courville, and Yoshua Bengio.
 \textit{Deep Learning}, MIT Press, 2016. http://www.deeplearningbook.org.

\bibitem[5]{Hinton12}Geoffrey Hinton, Li Deng, Dong Yu, George Dahl, Abdel-rahman Mohamed, Navdeep Jaitly, Andrew Senior et al. \textit{Deep neural networks for acoustic modeling
in speech recognition}, Signal Processing Magazine 29 (2012), 82\textendash 97.

\bibitem[6]{Krizhevsky12}Alex Krizhevsky, Ilya Sutskever, and Geoffrey E. Hinton. \textit{Imagenet classification with deep convolutional neural
networks}, Advances in Neural Information Processing Systems 25 (2012),
1097\textendash 1105.

\bibitem[7]{LeCun15}Yann LeCun, Yoshua Bengio, and Geoffrey Hinton. \textit{Deep
learning}, Nature 521 (2015), 436\textendash 444. 

\bibitem[8]{Cybenko89}George Cybenko. \textit{Approximation by superpositions of a sigmoidal function}, Mathematics of control, signals and systems 2, no. 4 (1989): 303-314.

\bibitem[9]{Hornik90}Kurt Hornik, Maxwell Stinchcombe, and Halbert White. \textit{Universal approximation of an unknown mapping and its derivatives using multilayer feedforward networks}, Neural networks 3, no. 5 (1990): 551-560.

\bibitem[10]{Hinton86}David E. Rumelhart, Geoffrey E. Hinton, and Ronald J. Williams. \textit{Learning representations by back-propagating errors},
Nature (1986), 323, 533-536.

\bibitem[11]{Labordere17}Pierre Henry-Labordere. \textit{Deep Primal-Dual Algorithm for BSDEs: 
Applications of Machine Learning to CVA and IM}, (November 15, 2017). Available at SSRN: https://ssrn.com/abstract=3071506 or http://dx.doi.org/10.2139/ssrn.3071506  

\bibitem[12]{Qi18}Haojie Wang, Han Chen, Agus Sudjianto, Richard Liu, and Qi Shen.
\textit{Deep Learning-Based BSDE Solver for Libor Market Model with Application to Bermudan Swaption Pricing and Hedging},  arXiv:1807.06622.

\bibitem[13]{Raissi18}Maziar Raissi. \textit{Forward-Backward Stochastic Neural Networks: 
Deep Learning of High-dimensional Partial Differential Equations},  arXiv:1804.07010.

\bibitem[14]{Brigo13}Damiano Brigo, Massimo Morini, and Andrea Pallavicini \textit{Counterparty credit risk, collateral and funding: with pricing cases for all asset classes}, John Wiley \& Sons, Mar 5, 2013.

\bibitem[15]{Gregory15}Jon Gregory. \textit{The xVA Challenge: Counterparty Credit Risk, Funding, Collateral and Capital}, John Wiley \& Sons, Oct 26, 2015. 

\bibitem[16]{Karoui97}N. El Karoui, S. G. Peng and M. C. Quenez. \textit{Backward
Stochastic Differential Equations in Finance}, Mathematical Finance,
Vol. 7, No. 1 (January 1997), 1-71.

\bibitem[17]{Pardoux90}E. Pardoux and S. G. Peng. \textit{Adapted
solution of a backward stochastic differential equation}, Systems
\& Control Letters 14 (1990) 55-61.

\bibitem[18]{Pardoux92}E. Pardoux and S. G. Peng. \textit{Backward
stochastic differential equations and quasilinear parabolic partial
differential equations}, Stochastic Partial Differential Equations
and Their Applications pp 200-217.

\bibitem[19]{Kingma14}Diederik P. Kingma, and Jimmy Ba.  \textit{Adam: A method for stochastic optimization}, arXiv:1412.6980. 2014 Dec 22.

\bibitem[20]{AndersenPiterbarg}Leif B. G. Andersen and Vladimir V. Piterbarg. \textit{Interest Rate Modeling. Volume 2: Term Structure Models}, Atlantic Financial Press, 2010.

\bibitem[21]{LeCun89}Yann LeCun, Bernhard Boser, John S. Denker, Donnie Henderson, Richard E. Howard, Wayne Hubbard, and Lawrence D. Jackel. 
\textit{Backpropagation applied to handwritten zip code recognition}, Neural computation 1.4 (1989), 541-551.

\bibitem[22]{Goodfellow14}Ian Goodfellow, Jean Pouget-Abadie, Mehdi Mirza, Bing Xu, David Warde-Farley, Sherjil Ozair, Aaron Courville, and Yoshua Bengio.
 \textit{Generative adversarial nets}. In Advances in neural information processing systems 2014 (pp. 2672-2680).

\end{thebibliography}
\end{document}